\documentclass[a4paper,oneside]{article}
\usepackage{graphicx}
\usepackage{multicol}

\newcommand{\be}{\begin{equation}}
\newcommand{\ee}{\end{equation}}
\newcommand{\ba}{\begin{array}}
\newcommand{\ea}{\end{array}}

\newcommand{\eV}{\,{\rm eV}}

\def\Red  {}

\def\Black{}
\def\Green{} % PANTONE 323
\def\Blue {}
\newcommand{\eq}[1]{~(\ref{eq:#1})}
\newcommand{\MeV}{\,{\rm MeV}}
\newcommand{\NP}{Nucl. Phys.}
\newcommand{\PRL}{Phys. Rev. Lett.}
\newcommand{\PL}{Phys. Lett.}
\newcommand{\PR}{Phys. Rev.}

\newcommand{\fig}[1]{~{\rm \ref{fig:#1}}}

\newcommand{\km}{\,\hbox{km}}
 
\newcommand{\md}[1]{\langle #1 \rangle}

\def\circa#1{\,\raise.3ex\hbox{$#1$\kern-.75em\lower1ex\hbox{$\sim$}}\,}
\makeatletter
\def\art{\@ifnextchar[{\eart}{\oart}}
\def\eart[#1]#2#3#4#5#6{{\rm #2}, {\em #3 \bf #4} {\rm (#6) #5} ({\em #1})}
\def\hepart[#1]#2{{\rm #2, \em#1}}
\newcommand{\oart}[5]{{\rm #1}, {\em #2 \bf #3} {\rm (#5) #4}}

\newcounter{alphaequation}[equation]
\def\thealphaequation{\theequation\hbox to
0.6em{\hfil\alph{alphaequation}\hfil}}
\def\eqnsystem#1{
\def\@eqnnum{{\rm (\thealphaequation)}}
\def\@@eqncr{\let\@tempa\relax \ifcase\@eqcnt \def\@tempa{& & &} \or
  \def\@tempa{& &}\or \def\@tempa{&}\fi\@tempa
  \if@eqnsw\@eqnnum\refstepcounter{alphaequation}\fi
\global\@eqnswtrue\global\@eqcnt=0\cr}
\refstepcounter{equation} \let\@currentlabel\theequation \def\@tempb{#1}
\ifx\@tempb\empty\else\label{#1}\fi
\refstepcounter{alphaequation}
\let\@currentlabel\thealphaequation
\global\@eqnswtrue\global\@eqcnt=0 \tabskip\@centering\let\\=\@eqncr
$$\halign to \displaywidth\bgroup \@eqnsel\hskip\@centering
$\displaystyle\tabskip\z@{##}$&\global\@eqcnt\@ne
\hskip2\arraycolsep\hfil${##}$\hfil& \global\@eqcnt\tw@\hskip2\arraycolsep
$\displaystyle\tabskip\z@{##}$\hfil
\tabskip\@centering&\llap{##}\tabskip\z@\cr}
\def\endeqnsystem{\@@eqncr\egroup$$\global\@ignoretrue} \makeatother

\begin{document}\twocolumn[
\centerline{hep-ph/0109172 \hfill CERN--TH/2001--220\hfill IFUP--TH/2001--22}
\vspace{5mm}
\Black
\vspace{0.5cm}
\centerline{\LARGE\bf\Red Which solar neutrino experiment}\vskip2mm
\centerline{\LARGE\bf\Red after KamLAND and Borexino?}
\medskip\bigskip\Black
  \centerline{\large\bf Alessandro Strumia$^\dagger$}\vspace{0.2cm}
  \centerline{\em Theoretical Physics Division,
CERN, CH-1211 Geneva 23, Switzerland }\vspace{0.4cm}
  \centerline{\large\bf  Francesco Vissani}\vspace{0.2cm}
  \centerline{\em INFN, Laboratori Nazionali del Gran Sasso,
Theory Group, I-67010 Assergi (AQ), Italy}

\vspace{1cm}
\Blue\centerline{\large\bf Abstract}
\begin{quote}\large\indent
We estimate how well we will know the
parameters of solar neutrino oscillations
after KamLAND and Borexino.
The expected error on $\Delta m^2$
is few per-mille in the VO and QVO regions,
few per-cent in the LMA region, and
around $10\%$ in the LOW region.
The expected error on $\sin^22\theta$ is around $5\%$.
KamLAND and Borexino will tell unambiguously which
specific new measurement,
dedicated to $pp$ solar neutrinos,
is able to contribute to the determination
of  $\theta$ and perhaps of $\Delta m^2$.
The present data suggest as more likely outcomes:
no measurement,
or the total $pp$ rate,
or its day/night variation.

% 
% We estimate how well we will know the
% parameters of solar neutrino oscillations after KamLAND and Borexino.
% The expected error on $\Delta m^2$ 
% is few per-mille in the VO and QVO regions,
% few per-cent in the LMA region, and 
% around $10\%$ in the LOW region.
% The expected error on $\sin^22\theta$ is around $5\%$.
% KamLAND and Borexino will tell unambiguously which
% specific new
% measurement dedicated to sub-MeV solar neutrinos
% could eventually be done with the minimal accuracy 
% necessary to give additional 
% information on  $\theta$ and maybe on $\Delta m^2$.
% The present data suggest that it could be: 
% no one, or the total $pp$ rate, or its day/night variation.

\Black
\end{quote}
\vspace{5mm}]

\footnotetext[2]{On leave from dipartimento di Fisica
dell'Universit\`a di Pisa and INFN.}

% \begin{figure*}[t]
% \vspace{-1cm}
% $$\hspace{-1mm}
% \includegraphics[width=60mm]{global.eps}~
% \includegraphics[width=60mm]{noClbench.eps}~
% \includegraphics[width=60mm]{noSSMbench.eps} $$
%   \caption[]{\em Present global fits at $90$ and $99\%$ CL of solar data in the $(\tan^2\theta,\Delta m^2/\eV^2)$ plane:
% (a) standard; (b) omitting the uncalibrated Chlorine rate;
% (c) the ``solar model independent fit'' described in~\cite{SSMindep}.\label{fig:today}}
% \end{figure*}

\section{Introduction}
The solar neutrino anomaly revealed in
  Homestake~\cite{ClSun},
Kam\-iokande~\cite{KaSun}, Gallex~\cite{GaSun} and SAGE~\cite{SAGEsun}
  has motivated the upgrades SuperKamiokande (SK)~\cite{SKsun}, GNO~\cite{GNOsun},
  and a new generation of experiments: SNO~\cite{SNOsun},
KamLAND~\cite{KamLAND} and Borexino~\cite{Borexino}.
  In the longer term, there are plans to attempt the
real-time detection of $pp$ 
neutrinos, thus covering
  the whole solar neutrino
spectrum~\cite{EsperimentiSubMeV}.
Depending on the choice of experimental technique,
it is hoped that
future sub-MeV experiments will be able to measure some of the following
properties of the solar neutrino flux at sub-MeV energies
\begin{itemize}
\item total rate;
\item day/night variations;
\item ``seasonal''  variations;
\item energy spectrum;
\item total rate in neutral and charge-current (NC and CC) reactions.
\end{itemize}
Today there are few disjoint best-fit 
solutions (usually na\-med LMA, LOW, VO, \ldots) and
these measurements could identify the true one.
In fact,
the survival probabilities $P_{ee}(E_\nu)$
for the present best-fit oscillations are not much different at $E_\nu\sim 10\MeV$
where we  have more experimental data,
but are significantly different at lower $E_\nu$.

  However, these
sub-MeV experiments will presumably start after
  SNO, KamLAND and Borexino
 have already identified the true solar-neutrino solution
and determined the solar-neutrino oscillation parameters.
  In this case, one should change the perspective and
  evaluate the potential of new experiments to
{\em improve} on the measurement of solar
oscillation parameters --- to be contrasted
with ``to {\em prove} the occurrence of oscillations''. From this
point of view, we answer the question in the title
by determining how well near-future
experiments are expected to contribute to these measurements.
This fixes the minimal necessary accuracy of new sub-MeV experiments.\footnote{
We do not consider other possible reasons why
one could be interested in sub-MeV solar neutrino experiments.
They can be used to verify 
(or contradict) existing results.
They could set bounds on exotic solutions
of the solar anomaly
(such as $\nu_e$ transitions into
sterile neutrinos, into extra-dimensional neutrinos,
into anti-neutrinos) or on
neutrinos with exotic properties (such as
a monster decay rate, 
or FCNC interactions, or magnetic moment, or else~\cite{v21}).
They can check the existence of the MSW effect and
the oscillation pattern fixed by  more precise experiments,
or to test solar model predictions,
detecting possible short-scale time variations.
However, helioseismology already provides accurate experimental information
on the static properties of the sun.
Finally, they could be used to demonstrate the validity of a new 
experimental method or technique.}

\begin{figure*}[p]\vspace{-0.5cm}
~\hfill $90\%$ CL \hfill\hfill $99\%$ CL ~\hfill~\\[-5mm]
$$
\includegraphics[width=85mm,height=21cm]{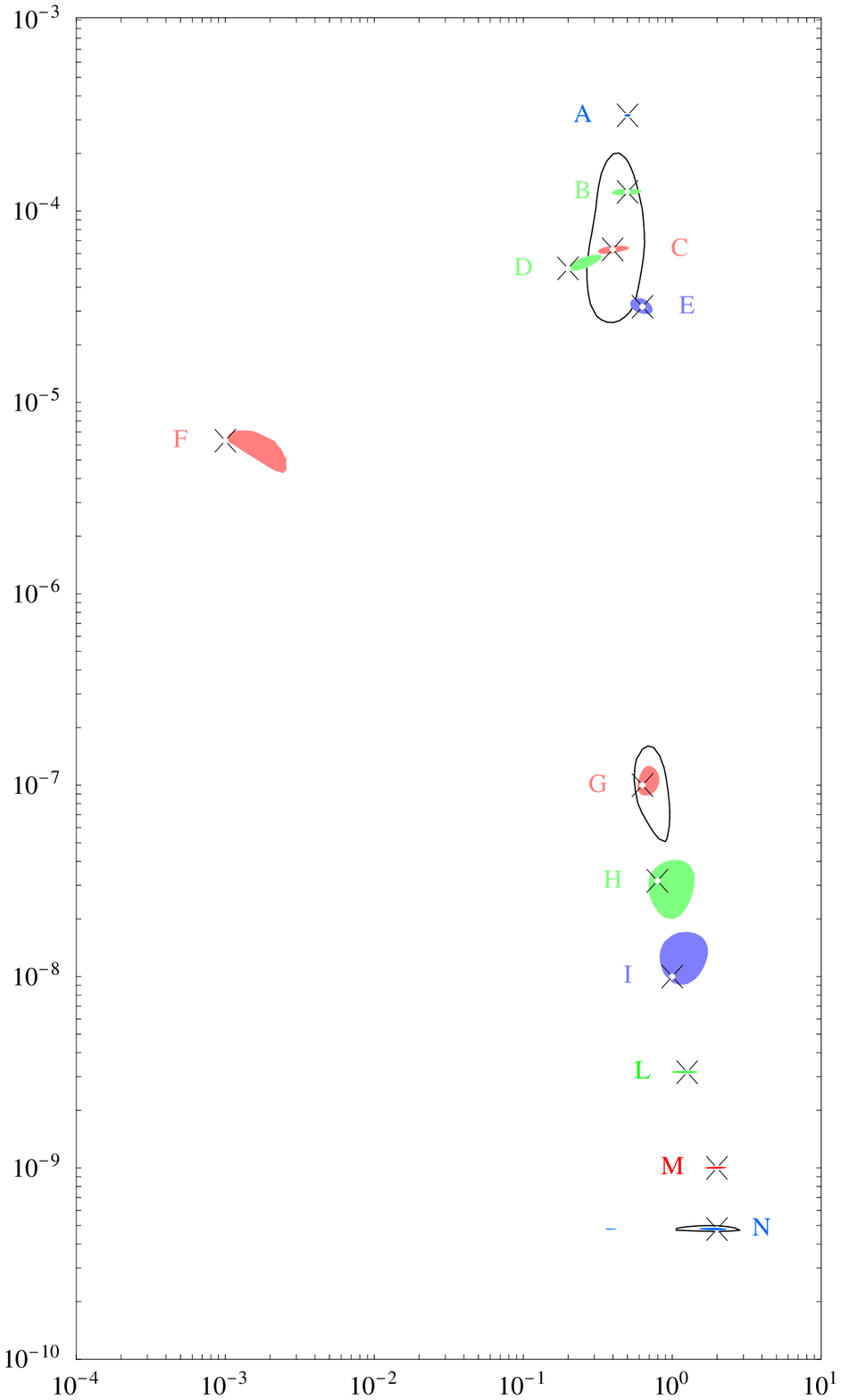}~
\includegraphics[width=85mm,height=21cm]{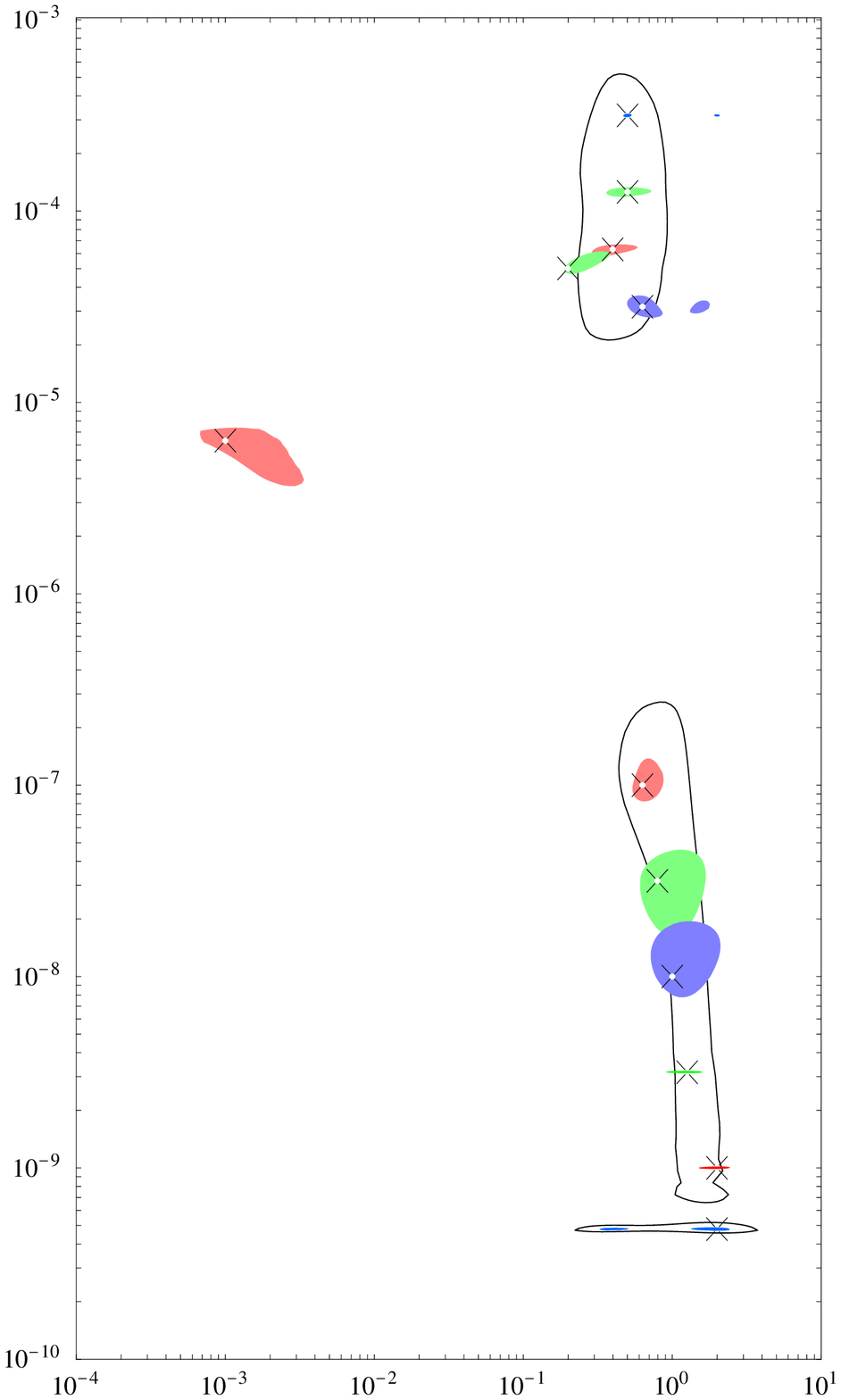}$$\vspace{-1cm}
  \caption[]{\em The continuous lines show the present global fit of solar data in the $(\tan^2\theta,\Delta m^2/\eV^2)$ plane.
The coloured regions are the simulated near-future fits (after KamLAND and Borexino)
for few benchmark points, marked as $\times$ and listed in table~1.
The left (right) plot shows fits at $90\%$ ($99\%$) confidence level.
The true oscillation solution is identified in all cases (LMA, SMA, LOW, QVO, VO).
Some simulated fits give very small regions: their main features 
can be read from table~1.
\label{fig:future}}
\end{figure*}

The conclusions (see table~1 or fig.\fig{future}) crucially depend
on the true value of the oscillation parameters $\Delta m^2$ and $\theta$.
A sub-MeV detector able to do different measurements but 
only with modest accuracy is never relevant 
to the measurement  of oscillations parameters.
Conversely, certain specific sub-MeV measurement could 
be relevant, if done precisely
enough. Near-future experiments are able to 
cover fairly  well all the possible cases
(often measuring $\Delta m^2$ with great precision), 
and will indicate unambiguously which is the remaining
relevant sub-MeV measurement.

\begin{table*}[t]
$$\begin{array}{|clcc|c|cccc|c|}\hline
\multicolumn{4}{|c|}{\hbox{\Blue Benchmark points\Black}} & \hbox{Present}&\multicolumn{4}{|c|}{\hbox{\Green Estimated near-future
uncertainty\Black}} &\hbox{\Red Useful sub-MeV\Black}\\ &\Blue \hbox{region} & \Delta m^2/\eV^2& \tan^2\theta\Black &\chi^2 - \chi^2_{\rm
best} &
\Green \hbox{main experiment} & \delta\Delta m^2 &
\delta\tan^2\theta& \delta P_{ee}\Black &\hbox{\Red measurement\Black }\\
\hline
\hbox{A}&\hbox{EI} &10^{-3.5} & 10^{-0.3} & 7.1 & \hbox{sub-KamLAND}^{\phantom{X}^{\phantom{~}}}
& 0.5\%   & 3\%  & 0.005  & \hbox{---}\\
\hbox{B}&\hbox{LMA}        & 10^{-3.9} & 10^{-0.3} & 3.5 & \hbox{KamLAND}         & 2\%     & 10\%  & 0.015  & \hbox{rate?}\\
\hbox{C}&\hbox{LMA$\cdot$} & 10^{-4.2} & 10^{-0.4} & 0 & \hbox{KamLAND}           & 2.5\%   & 10\%  & 0.015  & \hbox{rate?}\\
\hbox{D}&\hbox{LMA}        & 10^{-4.3} & 10^{-0.7} & 15.6 & \hbox{KamLAND}    & 4\%     & 10\%  & 0.02  & \hbox{rate?}\\
\hbox{E}&\hbox{LMA}        & 10^{-4.5} & 10^{-0.2} & 6.7  & \hbox{KamLAND}        & 5\%     & 10\%  & 0.01  & \hbox{rate?}\\
\hbox{F}&\hbox{SMA}        & 10^{-5.2} & 10^{-3.0} & 21 & \hbox{SK, SNO, Borexino}& 15\%    & 20\%  & 0     & \hbox{spectrum!}\\
\hbox{G}&\hbox{LOW$\cdot$}&10^{-7.0} & 10^{-0.2} & 3.4 & \hbox{Borexino day/night}& 10\%    & 10\%  & 0.01  & \hbox{rate, day/night?}\\
\hbox{H}&\hbox{LOW}       &10^{-7.5} & 10^{-0.1} & 8.2 & \hbox{Borexino day/night}& 20\%    & 20\%  & 0.03  & \hbox{rate, day/night}\\
\hbox{I}&\hbox{border}\!\!\!\!\! &10^{-8.0} & 1 & 9.1 & \hbox{KamLAND?}           & 20\%    & 20\%  & 0.03  & \hbox{rate, day/night!} \\
\hbox{L}&\hbox{QVO}       &10^{-8.5} & 10^{0.1} &8.0 & \hbox{Borexino seasonal}   & 0.4\%   & 10\%  & 0.015  & \hbox{rate}\\
\hbox{M}&\hbox{QVO}       &10^{-9.0} & 10^{0.3} &10.0&\hbox{Borexino seasonal}& 0.5\%   & 10\%  & 0.015  & \hbox{rate}\\
\hbox{N}&\hbox{VO}\cdot&10^{-9.32} & 10^{0.3} & 4.2 & \hbox{Borexino seasonal}& 0.7\%   & 10\%  & 0.015  & \hbox{rate}\\
\hbox{O}&\multicolumn{2}{l}{\hbox{no oscillations}} &-&
%% \hbox{NO} &0 &0 &
 50 &\hbox{SK, SNO, Borexino} & -&- & -&\hbox{rate!}\\
\hline
\end{array}$$
\caption{\em A few benchmark 
points (ordered according to the value of $\Delta m^2$), 
their present status,
their near-future status, and the most relevant 
measurement left for sub-MeV experiments.
$\delta P_{ee}$ is the error on the 
average survival probability of $pp$ neutrinos.
In absence of oscillations, near future 
experiments will indirectly 
measure $P_{ee}$ with $\pm0.06$ error.}
\end{table*}

\section{The near-future situation}
We focus mainly on the oscillations of three active neutrinos,
assuming that atmospheric oscillations do not 
affect solar neutrinos (in the standard notation, this amounts
to have a large $\Delta m^2_{\rm atm}$ and a small $\theta_{13}$)
because the relevant experiments~\cite{ClSun,KaSun,GaSun,SAGEsun,SKsun,
GNOsun,SNOsun,SKatm,Macro,Bugey,CHOOZ,Karmen} (except LSND~\cite{LSND}, see
page~\pageref{sterile}) indicate that this is the physically relevant case.
A three flavour analysis of
solar (and atmospheric) data is no longer relevant~\cite{kx23}:
reactor experiments~\cite{CHOOZ} directly demand a
small $\theta_{13}<15^\circ$ at $95\%$ CL, so that
solar oscillations are presently determined
by the usual two parameters,     $\Delta m^2$
and $\theta.$\footnote{$\theta_{13}$ could
be large enough to give detectable effect
in future solar neutrino experiments.
But we will not consider this possibility in our analysis,
because in this case $\theta_{13}$ will be
measured much more precisely by future long-baseline experiments.}

The continuous lines in fig.~\ref{fig:future}
show the present global fit~\cite{standardFit,statistica}.
More or less acceptable fits can be obtained for
a wide range of $\Delta m^2$ and for a
large mixing angle.
The best-fit  solutions have
$\chi^2_{\rm best}\approx 33$ from 41
experimental inputs and 2 free parameters;
they lie in the LMA region, and have a
$\Delta m^2$ around $10^{-4}\eV^2$.
The other solutions with smaller $\Delta m^2$
(named LOW, QVO, VO) are not significantly worse:
they poorly fit the data where a `solar anomaly' is present,
i.e.\ the total rates, but satisfy well the bounds
from data consistent with no oscillations,
i.e.\ spectral distortions, seasonal and day/night variations.
However, the discrimination is not sharp: e.g.,
if GNO should decrease the central value of 
the Gallium counting rate down to $65\pm 5$ SNU, 
($\sim 2 \sigma$ below the present value),
the global fit would favor the LOW solution. 
%%%%%%%%%%%%%%%%%%%%%%%%%%%%%%%%%%%%%%%%%%%%%%%%%%%%%%%%%%%
%{\em If GNO will find $(65\pm 3)$ SNU, 
%the  LOW solution will be half-disfavoured than today. }
%%%%%%%%%%%%%%%%%%%%%%%%%%%%%%%%%%%%%%%%%%%%%%%%%%%%%%%%%%%

In order to simplify the discussion
it is useful to focus on few benchmark points
that span the qualitatively different, still allowed solutions.
These points (denoted as point A, B, C, ...)
are listed in table~1, and drawn
as `$\times$' in fig.~\fig{future}.
Some of them have a high
$\Delta\chi^2\equiv \chi^2 - \chi^2_{\rm best}$
(i.e.\ are significantly disfavoured, roughly at
$(\Delta \chi^2)^{1/2}$ standard deviations: for
example there is a $\sim 7\sigma$ evidence for LMA versus no oscillations).
We retained them in order to have
a conservative sampling of all possible cases.
The points D, M, N, O have a significantly {\em lower}
$\Delta\chi^2$
in a solar-model-independent analysis,
where the solar-model predictions for the Boron
and Beryllium fluxes are not used
(see~\cite{SSMindep} for a precise description).

Fig.\fig{future} and table~1
illustrate the expected
near-future achievements.\footnote{When the
uncertainty is non Gaussian,
one standard deviation errors have been
replaced by one half of two standard deviation errors,
if this gives a more conservative result.
The $90\%$ and $99\%$ CL contours in our fits in fig.\fig{future}
actually correspond to $\Delta\chi^2 =4.6$ and $9.2$,
as obtained for two unknown parameters ($\Delta m^2$ and $\theta$)
using the Gaussian approximation, that is reasonably
accurate~\cite{statistica}.}
In the last column of table~1
we summarize which sub-MeV measurements
could provide us with further, useful information.
When a certain sub-MeV measurement  is
crucial (not very interesting), we mark it
by a `!' (`?').
While $\Delta m^2$ can  be often measured very accurately,
the determinations of $\theta$
may be less reliable. Indeed, the results on $\theta$
depend strongly on solar-model predictions; they
could change e.g.\ if the Beryllium flux were lower
than its predicted value.
For points outside the LMA region,
the error on $\theta$ (and consequently on $P_{ee}$)
can be accurately estimated only after knowing
the true results of near-future experiments,
and could differ by a factor 2 from
the values quoted in table~1.

The values of $\delta P_{ee}$ in the penultimate column
of table~1 are the 1 standard-deviation,
near-future uncertainty on the survival probability
of the total $pp$ rate, as detected using
$\nu_e e\to \nu'_e e'$ scattering, with electron
kinetic energy  larger than $T_{e'}>0.1\MeV$
(taking into account the smaller $\nu_{\mu,\tau}$ NC cross section,
the suppression in the total rate due to oscillations
is roughly given by $0.8\ P_{ee}+0.2$).
This quantity is important for 
sub-MeV experiments~\cite{EsperimentiSubMeV},
since it informs us on how accurately those based 
on elastic scattering (as {\sc Heron, 
Clean, Xmass, Genius, Hellaz}, etc.) 
should measure the $pp$ rate.
Within the uncertainties,
$\delta P_{ee}$ is also relevant for
experiments based on inverse $\beta$-decay
(as {\sc Lens, Moon,} etc.).

One should recall that existing Gallium experiments
are sensitive to $pp$ neutrinos
(above $0.24\MeV$), that would induce half of their neutrino events
in absence of oscillations.
In the present perspective, we are lead to study
how well the true rate of $pp$ neutrinos can
be reconstructed from the total Gallium rate,
after subtracting the values of the other fluxes (Boron, Beryllium, \ldots)
$$R^{pp}_{\rm Ga} = R^{\rm total}_{\rm Ga}-R_{\rm Ga}^{\rm others}$$
as measured by the near-future (and present) experiments.
Concerning the Boron flux, SK and SNO
find that, within their accuracy, the
survival probability of Boron neutrinos is energy-independent.
Therefore, we directly use the
Boron rate measured at SNO with $8.5\%$ error.
The error on the Boron contribution to the
Gallium rate is $\pm 1.5$ SNU, dominated by the $30\%$ uncertainty
on its Gallium cross
section~\cite{LisiChiq}.

Concerning the Beryllium and CNO fluxes,
Borexino (and, maybe, KamLAND) should measure them.
We assume that the error on Beryllium and CNO
contributions to the total Gallium rate will again be dominated
by the $\sim 10\%$ uncertainties
on their Gallium cross section, so that the total
error on $R_{\rm Ga}^{\rm others}$ will be $\pm 2$ SNU.
We neglect the $pep$, $hep$ and F contributions,
that are smaller than this error.
The experimental error on $R^{\rm total}_{\rm Ga}$ is
today $\pm 5$ SNU, and GNO should lower
it down to $\pm (3\div 4)$ SNU.
This means that the $pp$ rate will be known within
$\pm (10\div 15)\%$ uncertainty, depending on its actual value
(here, we assumed a large mixing angle).
In this example, the average value of the survival
probability for $pp$ neutrinos (around $0.5$)
will be known with a $\pm0.06$ uncertainty.

In the rest of the paper we
motivate and discuss in detail the
results summarized in table~1 and fig.\fig{future}.

\subsection*{LMA and EI}
The present data favour the LMA region, 
and do not significantly disfavour
energy independent (EI) solar oscillations, obtained for 
$\Delta m^2 \circa{>} 2~10^{-4}\eV^2$~\cite{Large12}.
This region should be fully covered by 
the KamLAND experiment~\cite{KamLAND},
which will detect reactor $\bar{\nu}_e$ using the reaction
$\bar{\nu}_e p\to e^+n$.
The accuracy of KamLAND has been discussed 
in~\cite{Barger,SSMindep,Murayama,GP} and more
importantly in~\cite{KamLAND}.
As in~\cite{KamLAND} we assume 
(too conservatively?) that 
KamLAND will use a cut on visible energy
$E_{e^+} + m_e > 2.6\MeV$ (where $E_{e^+} = E_\nu - m_n+m_p$)
in order to avoid the background due to ambient $\bar{\nu}_e$.
We assume a rate of
550 events per year without oscillations,
a $2\%$ overall uncertainty on the 
reactor flux and no background.
The regions in fig.\fig{future} 
surrounding the points B, C, D, E
show how well KamLAND is expected to
determine the oscillation 
parameters $\Delta m^2$ and $\theta,$
after three years of data-taking.
The value of $\Delta m^2$ can be measured accurately because 
the initial spectrum is well known, and
the energy resolution is sufficient to show the 
first oscillation dip.

However, if $\Delta m^2 \circa{>} 2~10^{-4}\eV^2$
(the precise value depends on the energy resolution)
KamLAND will only see averaged oscillations;
thence, it will be unable to
measure $\Delta m^2$ with good sensitivity~\cite{SSMindep}.
%In this case, the chances to estimate the oscillation parameters  
%will be meager;  long-baseline experiments (even those 
%using a neutrino factory beam) will be unable 
%to measure $\Delta m^2$  {\em accurately}
%(see the discussion in~\cite{SSMindep}).
Therefore, for point~A we considered 
a new reactor experiment, with a baseline of $20\km,$ 
a rate of $3000$ events/year above $E_{e^+}+m_e > 2.6\MeV,$
and used three years of data for the estimate.
This experiment has been named `sub-KamLAND' in table~1
and in the following, 
because it is {\em less} demanding than KamLAND. 
  \footnote{Here one crucial question is: if $\Delta m^2 \circa{>}
  2~10^{-4}\eV^2$, do we need  such a reactor experiment, or
  long-baseline experiments will do a better job?
  The answer is that even using a neutrino factory beam
  long-baseline experiments  will be unable
  to measure the solar parameters {\em accurately}:
  see~\cite{SSMindep} for a discussion.
The main reason is that $\Delta m^2_{12}$ effects are entangled with effects due to $\theta_{13}$.
To disentangle them one need to compare some combination of observables,
different from the one that gives the highest rate
($\nu_e\to \nu_\mu$ at a baseline of $L\sim 700\km$).
Detailed studies of neutrino-factory capabilities
  assumed that the solar parameters
  will be accurately measured by other experiments~\cite{nuf}.}

% \footnote{
% In fact, in a reasonably good approximation (see~\cite{SSMindep} for details)
% the rates in long-baseline observables are proportional to
% $|\Delta_{e\mu}|^2$, where
% $\Delta_{e\mu} \approx [\theta_{13}\Delta m^2_{\rm atm}+
% \frac{1}{2}e^{i\phi}\Delta m^2_{\rm sun}]/\sqrt{2}$ 
% is the $e\mu$ element of the neutrino squared
% mass matrix.
% Long-baseline experiments can measure $|\Delta_{e\mu}|$ accurately.
% However, only a combination of different observables
% can distinguish the `solar' and `atmospheric' contributions to $\Delta_{e\mu}$:
% some of these observables need an oscillation channel, baseline or beam energy
% different from the one that gives the highest rate
% (e.g.\ one can hope in a non zero CP-violating phase, $\phi\neq 0$).
% Since rates will be a major limiting factor of neutrino beam experiments,
% one can understand why a reactor experiment can determine the
% solar parameters much more accuarely.
% We do not find useful to present fits of simulated data,
% because neutrino-factory observables depend on too many unknown parameters
% (e.g.\ all oscillation parameters).
% % a detailed discussion  would be a long list of ``if\ldots then \ldots else''.
% As a matter of fact, detailed studies of neutrino-factory capabilities
% assumed that the solar parameters
% will be accurately measured by other experiments~\cite{nuf}.}

Despite the assumed $2\%$ uncertainty 
on the initial $\bar{\nu}_e$ flux,
$P_{ee}$ (and consequently the mixing 
angle $\theta$) can be measured with an error less than $2\%$,
{\em if} an oscillation signal is seen.
In fact, the accuracy in the determination 
of $P_{ee}$ depends strongly on its value, 
being maximal when $\theta\sim \pi/4.$
It is possible to detect small deviations from maximal mixing.
This can be understood in a qualitative 
way by considering an energy bin around the 
first oscillation dip, where $P_{ee}\sim 0$.
It will contain a number of 
events $N_{\rm obs}$ much smaller than the no 
oscillation prediction, $N_{\rm O}$.
Having assumed a negligible background,
the survival probability in that bin 
can be measured with error $\delta P_{ee} \sim\sqrt{N_{\rm obs}}/N_{\rm O}$.
This feature is more pronounced at the 
hypothetical sub-KamLAND than at KamLAND,
where the $\bar{\nu}_e$ are produced 
by various reactors with different path-lengths,
so that the first oscillation dip is partially averaged.
(The situation would improve if KamLAND 
could reconstruct the direction of the neutrinos).

% We assume that directionality of the neutrons is not used.

\medskip

We now discuss how sub-MeV experiments 
could improve on this situation.
Solar neutrinos with energy
\begin{equation}\label{eq:noMSW}
E_\nu \circa{<} \frac{\Delta m^2}{2\sqrt{2}G_{\rm F} N_e^\odot} 
\approx 1\MeV \frac{\Delta m^2}{10^{-5}\eV^2}
\end{equation}
(where $N_e^\odot$ is the electron density 
around the region of neutrino production)
do not experience the MSW resonance in the sun. 
Therefore, their oscillation probability
is roughly given by averaged vacuum 
oscillations, $P_{ee}\approx 1-\frac{1}{2}\sin^22\theta$.
In first approximation, this implies
that sub-MeV experiments have nothing 
to tell about $\Delta m^2$, but
could give information on $\theta$. Note however that
KamLAND will fix the value of $P_{ee}$ with an error less than $\pm0.02,$
as shown in table~1.
Though the precision of sub-MeV experiments is 
ultimately limited by the $1\%$ solar model uncertainty on the $pp$ flux,
it seems unrealistic to aim at this level of accuracy, 
and even difficult to compete with KamLAND 
(or other upgraded reactor experiments).
On the other hand, with
a precise determination of $\Delta m^2$ 
and $\theta$ from a reactor experiment,
sub-MeV experiments could be used to finally 
test solar model predictions, as suggested long time 
ago~\cite{1964}.

Reactor experiments can only measure $\sin^2(2 \theta)$:
the discrimination between $\theta$ and $\pi/2-\theta $
has to be performed by relying on matter effects acting on solar neutrinos.
The present data prefer $\theta<\pi/4$, 
with a few standard deviations significance
(its precise value
depends on the actual value of $\theta$ and $\Delta m^2$~\cite{GP}).
Which new experiments are best suited for this issue?
Neutrino fluxes are better known 
at lower $E_\nu$, but in the LMA region matter effects
are larger at $E_\nu\sim 10\MeV$ than at $E_\nu \circa{<}\MeV$.
At larger $E_\nu,$ the uncertainty on the Boron flux can be circumvented
by the NC and CC measurements at SNO (and SK).
The NC/CC ratio today prefers $\theta<\pi/4$ at $\sim 1.5$ standard deviations,
and could provide a direct discrimination in the near future.
If $\Delta m^2$ is in the lower part of the LMA region,
earth matter effects at SK and SNO~\cite{LMAmatter} could discriminate
$\theta$ from $\pi/2-\theta $.
%     la differenza in A e'di 0.01--0.03 nei 3 punti in lower LMA.
Matter corrections to the spectrum of $\bar{\nu}_e$ from a future supernova 
should also give a clear discrimination:
unlike $\nu_e$, $\bar{\nu}_e$ cross 
a resonance if $\theta > \pi/4$.
% Actually, it has been argued that neutrinos from 
% SN1987A, already, disfavour the LMA regions 
% with $\theta>\pi/4,$ and those  
% with small $\Delta m^2$ (VO, QVO, \ldots)~\cite{SN1987A}.
%This statement is based on
%supernova models and on 
%Kamiokande and IMB data.
%Thirty  years ago, the solar anomaly was based on similarly unsafe
%theoretical and experimental  grounds.

\subsection*{LOW}
The most promising LOW signals in near-future experiments are
earth matter effects at Borexino and, maybe,
at KamLAND~\cite{KamLAND,Borexino-LOW}. In fact,
after having excluded the LMA region, KamLAND could
be converted into a solar neutrino experiment.
In absence of oscillations,
Borexino (KamLAND) is expected to
have a signal rate of 59 (280) events/day for
the central value of the
Beryllium, CNO, $pep$ fluxes predicted by~\cite{BP} and
a background rate of 19 (130)
events/day, known with $ \delta b =10\%$ ($50\%$\footnote{Here
we hope to be too pessimistic.
We assume that the fiducial volume
will be $60\%$ of KamLAND~\cite{KamLAND}; this is why
our rates are $60\%$ lower than those
employed in~\cite{Borexino-LOW,Borexino-QVO2}.}) error.
Even if the background were more uncertain,
the seasonal variation of the solar
neutrino flux (due to the excentricity of the earth orbit)
would allow a measurement of the signal
rate with an interesting accuracy~\cite{KamLAND,Borexino-LOW}.
We concentrate on Borexino and consider
only the signal due to the Beryllium line (46 events/day),
after three years of running.

At present, it is not clear
if it will be possible to measure the CNO
and $pep$ contributions,
using a sufficiently background-free energy region
above the Beryllium line.
A pessimistic attitude would require to assume that
it will be only possible to use those
events with recoil $e$ energy
between $0.25$ and $0.8\MeV$,
generated by Beryllium (46 events/day),
CNO (10 events/day), $pep$ (2 events day)
neutrinos~\cite{Borexino} plus of course the background.
However, we remark
that this possible limitation would be a problem to test
of solar models, but would have just a little
effect on the  determination of $\Delta m^2$ and $\theta$.

\setcounter{footnote}{0}

We perform our analysis along the lines of~\cite{Borexino-LOW},
but paying more attention to the determination of the oscillation parameters,
rather than to the discovery of oscillation signals.
We divide the simulated data into $N_{\rm bins} = 8+1$  zenith bins
(8 night zenith bins equally spaced in $\cos \theta_{\rm zenith}$,
plus one day bin) and define
\begin{eqnarray}\label{eq:chiqstandard}
\chi^2 &=& \sum_{i=1}^{\rm N_{\rm bins}}\frac{(N_i- S_i - b B_i)^2}{N_i}+\\
&&+
\bigg(\frac{\Phi_{^7{\rm Be}}-\Phi^{\rm BP}_{^7{\rm Be}}}{\delta
\Phi^{\rm BP}_{^7{\rm Be}}}\bigg)^2+
\frac{(b-1)^2}{\delta b^2},\nonumber
\end{eqnarray}
where $N_i$ is the measured number of events;
$S_i$ is the expected number of signal events
(proportional to $P_{ee}$ and to the solar flux $\Phi_{^7{\rm Be}}$);
$b B_i$ is the expected number of background events (the factor $b$
takes into account the overall uncertainty on the background rate).
Finally the $\chi^2$ in eq.\eq{chiqstandard}
is added to the $\chi^2$ from present data,
properly taking into account the correlation
of the theoretical uncertainties on
the Beryllium flux~\cite{BP,LisiChiq}.
A $\chi^2$ as in eq.\eq{chiqstandard}, when
minimized with respect to the `nuisance' unknown parameters
(here $b$ and $\Phi_{^7{\rm Be}}$; more
generically the solar-model parameters and
the detection cross sections)
gives the same $\chi^2$ defined in~\cite{LisiChiq},
where all uncertainties are summed in quadrature,
obtaining a big error matrix.

The simulated fit for point G (which gives the
best-fit in the LOW region) is similar to
the corresponding result in fig.~6 of~\cite{Borexino-LOW}
(where KamLAND instead of Borexino was considered).
The accuracy is worse at point H, because earth matter effects
diminish with $\Delta m^2$.
We do not show simulated fits for points located
around the highest $\Delta m^2$ values allowed in the LOW region
(that will be soon tested by SNO and are already 
directly disfavoured by the non-observation
of a day/night asymmetry at SK and Gallex/SAGE/GNO~\cite{GNOseasonal}): 
due to the
large earth matter effects, the accuracy on the determination of $\Delta m^2$
would be so good that one should carefully take into account
the uncertainty in the profile density of the earth
(while this is not an issue for the
points that we have selected).

The survival probability of sub-MeV neutrinos
is given by adiabatic conversion:
$P_{ee}=\sin^2 \theta$ during the day.
An accurate measurement of the $pp$ rate
would provide the safest determination of $\theta$,
because solar-model predictions will play little r\^ole.
It will be interesting to perform this measurement even if
present and near-future experiments
will nominally give a somewhat more
accurate determination of $\theta$.

Furthermore,
earth matter effects give a day/night variation of the $pp$ rate,
allowing to measure also $\Delta m^2$.
However, even if matter effects are larger at $pp$ energies
than at higher energy, it is
more convenient to study $\Delta m^2,\theta$
at Borexino (KamLAND, or new experiments based on inverse $\beta$-decay)
due to the monochromaticity of Beryllium neutrinos
and to the larger event rate.

\subsection*{LOW/QVO boundary}
We pragmatically define the boundary
between the LOW and QVO regions~\cite{QVO}
as
$$\Delta m^2({\rm LOW}) \ \circa{>}\ 10^{-8}\eV^2\ \circa{>} \
\Delta m^2({\rm QVO})$$
because this is the critical
$\Delta m^2$ under which  Borexino
(KamLAND) should observe anomalous
seasonal effects (see e.g.\ \cite{KamLAND,Borexino-QVO1,Borexino-QVO2}),
rather than the day/night effects characteristic of the
LOW region~(see e.g.~\cite{KamLAND,Borexino-LOW}).
The regions considered in the
rest of this paper will be soon disfavoured, if 
SNO finds a day/night asymmetry.

No unmistakable signal of solar neutrino
oscillations can be observed by near-future experiments
if $\Delta m^2$ lies around this critical value.
In view of this situation,  we tried to
exploit, {\em in this particular point},
the full capability of near-future experiments by
optimistically
assuming that KamLAND will be converted
into a solar neutrino experiment as described above.
KamLAND would detect a hint of day/night effect,
giving some information on $\Delta m^2$.
Borexino (and existing data) would provide
instead the dominant information on $\theta$.
By combining these pieces of information,
we obtain the simulated fit for point I.

Though the near-future
uncertainty in $\Delta m^2$
is significantly smaller than
the present uncertainty, it
remains rather big.
It should be understood that
the improvement is mainly due to
the fact that all other regions with
larger and smaller $\Delta m^2$ 
will be firmly excluded, because they
predict unobserved clear signals. 
To prove this, we omitted the only positive signal
(attributed to KamLAND) and obtained roughly
the same $\Delta m^2$ interval.

A measurement of the $pp$ rate could give additional
informations on $\theta$.
However a measurement of $\Delta m^2$
resulting from a global fit may be felt as unsatisfactory.
As discussed in the `no oscillations' section,
the detection of Beryllium neutrinos 
would ensure that
a solar neutrino anomaly is present,
but we would still not know if it is due to oscillations.
A sub-MeV experiment could discover
an unmistakable oscillation signal:
the neutrino energy at which earth matter
effects induce a maximal day/night asymmetry is
$$E_\nu^{\rm res} = \frac{\Delta m^2}{2\sqrt{2}G_{\rm F}
N_e^\oplus} \approx \frac{\MeV}{40} \frac{\Delta m^2}{10^{-8}\eV^2},$$
where $N_e^\oplus$ is the electron density of the earth mantle.
Because of the low value of $E_\nu^{\rm res}$,
one would need a big real-time detector,
with an energy threshold as low as possible
in order to detect this effect.

However, at a detector with a threshold $T_e>0.1$ MeV
(such as {\sc Heron, Clean, Xmass, Genius, Hellaz} etc.)
the day/night asymmetry at point I
(a $\sim 4$ \% excess in nighttime) is only twice
larger  than at Borexino or KamLAND, that could
instead have many more events.
A very low threshold, maybe as low as 11 keV,
could be attained by the {\sc Genius} experiment
with a rate of 18 $pp$ events/day
(assuming a mass of 10 ton).
However only few $\%$ of the $pp$ neutrinos 
have energy below $0.1\MeV,$ and they 
cannot be individually identified by 
the elastic scattering reaction.

\subsection*{QVO}
The distance between the earth and the sun varies as
$$L(t) = L_0[1- \epsilon \cos (2\pi t/{\rm yr}) + {\cal O}(\epsilon^2)]$$
where $L_0 = 1.496~10^8\km$ is the
astronomical unit, $\epsilon=0.0167$ is the
excentricity of the earth orbit, and $t$ is the
time since the perihelion (around $4{\rm th}$ of January).
This variation induces a modulation of the survival
probability in QVO and VO regions,
that can be investigated at Borexino, and eventually at KamLAND,
by means of the almost mono-energetic Beryllium neutrinos,
$E_{^7{\rm Be}}=0.863\MeV$~\cite{Borexino-QVO1,Borexino-QVO2}.

\begin{figure}[t]
$$
\includegraphics[width=80mm]{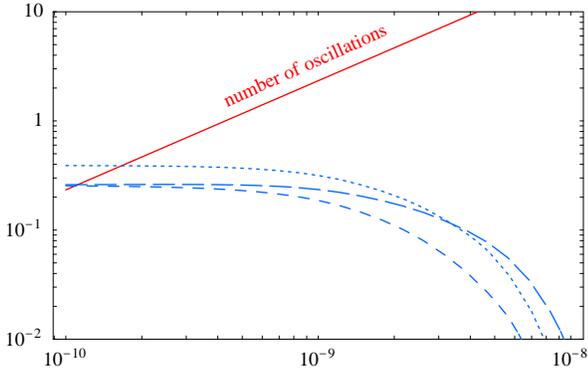} $$
  \caption[]{\em
Number $N_{\rm osc}$ and amplitude $a$ of seasonal oscillations
as function of $\Delta m^2$ in $\eV^2$ for
$\tan^2\theta = \{1/\sqrt{10},1,\sqrt{10}\}$ (long dashed, dotted, dashed).
\label{fig:seasonal}}
\end{figure}

The time variation of the
survival probability  is~\cite{QVO}
$$P_{ee}(t) = \langle P_{ee}\rangle+ \sin2\theta \sqrt{P_C(1-P_C)} D
\cos (kL(t)+ \delta),$$
where $k={\Delta m^2}/{2 E_{^7{\rm Be}}}$
and $\delta\sim 0.13\, k R_{\odot}$~\cite{QVO}.
The number of oscillations met in one semester
can be large (fig.\fig{seasonal}):
\begin{equation}\label{eq:Nosc}
N_{\rm osc} = \frac{\epsilon L_0
\Delta m^2}{2\pi E_{^7{\rm Be}}}\approx \frac{\Delta m^2}{0.85~10^{-9}\eV^2}.
\end{equation}
These oscillations get washed
for increasing $\Delta m^2,$ due to the
MSW effects inside the sun 
(taken into account by the crossing-probability factors $P_C$)
and to the finite width
of the Be line (taken into account by 
the factor $D$, given by the Fourier transform~\cite{QVO} of
the Beryllium `line' spectrum~\cite{BeSpectrum}),
as illustrated in fig.\fig{seasonal}.

We perform the simulated fits using a $\chi^2$ with
a large number of seasonal bins: $N_{\rm bins} \sim 10\cdot N_{\rm osc}.$
Of course, $N_{\rm osc}$ is a priori unknown in actual analyses,
and should be extracted from the data;
a definition of $\chi^2$ that avoids this problem, and
allows to exploit all the data is discussed in Appendix A.
Anyhow, our conclusion is that $\Delta m^2$ can be
measured with surprisingly good precision
(see table~1, or enlarge fig.\fig{future}).
The point is that
when the number of  oscillations $N_{\rm osc}$ is big,
Borexino acts as an interferometer.

\begin{figure}[t]
$$
\includegraphics[width=80mm]{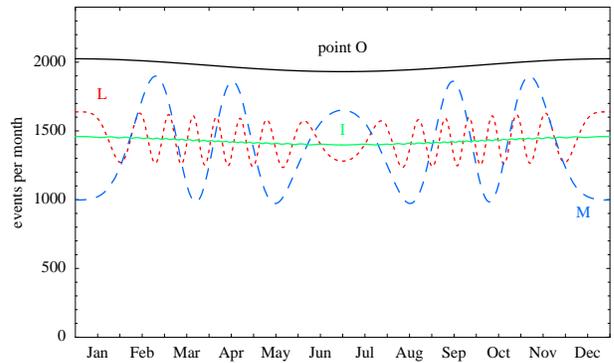} $$
  \caption[]{\em Seasonal variation of the signal at
Borexino for the benchmark points.
The upper line is the no-oscillation case.
\label{fig:seasonalBorexino}}
\end{figure}

These results can be understood by a simple analytical estimate.
First, we display in fig.\fig{seasonalBorexino}
the signal as a function of time,
for the benchmark points I, L and M.
The number of events at Borexino can be written as:
\begin{equation}
\label{eq:ka}N \propto 1+ a \cos (k L+\delta).
\end{equation}
considering the events due to Beryllium neutrinos (CC and NC)
and the estimated background,
and omitting the geometrical $1/L^2$ flux factor.
The amplitude of oscillations $a$ is
plotted in fig.\fig{seasonal},
for three choices of the mixing angles.
The average value of $N$ during the time periods with
$\cos (kL+\delta)>0$ ($<0)$ is $N_\pm \propto 1 \pm 2 a/\pi$.
Thence, one gets an ``asymmetry'' (systematic excess) of events:
$$\frac{4 a}{\pi} = \frac{N_+ - N_-}{(N_+   +
N_-)/2}\pm \sqrt{\frac{2}{N_+ + N_-}}.$$
The error on $a$ can be small even if
$N_{\rm osc}$ is big:
$$\delta a = \frac{\pi}{4} \sqrt{\frac{2}{N_++N_-}} =
0.005\sqrt{\frac{50000}{N_++N_-}}.$$
The `seasonal' oscillation is detected if the
amplitude $a$ is sufficiently larger than its uncertainty $\delta a$,
and this happens if
$\Delta m^2 \circa{<}0.5~10^{-8}\eV^2.$
Values of $\theta<\pi/4$ are more suitable;
see again fig.\fig{seasonal}.

Once a seasonal signal is seen
with $N_\sigma = a/\delta a$ standard-deviations,
the number of oscillations $N_{\rm osc}$ can be measured with an error
$\delta N_{\rm osc} \approx 1/2 N_\sigma$ that
does not depend on $N_{\rm osc}$.
Using eq.\eq{Nosc}, the consequent error on $\Delta m^2$ is
\begin{equation}\label{eq:dm}
\frac{\delta\, \Delta m^2}{ \Delta m^2}
= \frac{\delta N_{\rm osc}}{N_{\rm osc}}
=\frac{1}{2N_\sigma N_{\rm osc}}\qquad\hbox{if $N_\sigma > $ few}.
\end{equation}
For example, from fig.\fig{seasonal} we can read that in point L $N_{\rm osc} \sim 7$ and $N_\sigma\sim 20$.
This explains why in table~1 we claim
an extremely accurate measurement of $\Delta m^2$.
The analytical approximation tells how our numerical results should be rescaled
in order to extend our analysis to KamLAND or to a longer data-taking period.
For example, with 9 times more signal, $\delta a$ would be 3 times smaller,
giving a {\em per-mille} determination of $\Delta m^2$.
The accuracy is limited by statistics (and by the small excentricity of
the earth orbit).
No knowledge of the Beryllium spectrum or of other
solar-model dependent features is required.

The overall phase in $\cos (kL_0+\delta)$ depends on the total
earth-sun distance (rather than on its excentricity variation)
and is therefore more strongly dependent on $\Delta m^2$.
It contains  additional information: it
separates the allowed region of $\Delta m^2$
into $\delta N_{\rm osc}/2\epsilon\approx\hbox{few}$ separate thin islands~\cite{Borexino-QVO2}.
We do not include this information in our fits:
the error or $\Delta m^2$
is already so small that it would not even be
possible to see this sub-structure in fig.\fig{future}.

Beryllium neutrinos contribution is $\sim 30\%$ of
the total rate measured at SAGE and Gallex/GNO.
These experiments are further limited by
a modest time resolution, $\Delta t\sim \hbox{weeks},$
and thence their sensitivity to seasonal variations is
reduced. No such a signal has been found,
which implies weak bounds on the oscillation
parameters~\cite{seasonalGallex},
usually neglected in global analyses.

Subsequent $pp$ experiments cannot improve on
this determination of $\Delta m^2$.
A measurement of the $pp$ rate would instead give a useful
information on the mixing angle.
However, this information would have a certain
degree of solar-model dependence.
Indeed, in the QVO region the survival probability
$\md{P_{ee}}=\frac{1}{2}+(P_C-\frac{1}{2})\cos2\theta$
lies somewhere between vacuum oscillations
($\md{P_{ee}}=1-\frac{1}{2}\sin^2 2\theta$
for $P_C=\cos^2\theta$) and adiabatic oscillations
($\md{P_{ee}}=\sin^2\theta$ for $P_C=0$),
as controlled by the crossing-probability
$P_C = [e^{\gamma \cos^2\theta}-1]/[e^{\gamma} -1]$ where~\cite{Parke}
$$
\gamma=\frac{\pi \Delta m^2}{E_\nu
|d\ln N_e/dr|_{\rm res}}\approx \frac{\Delta m^2/E_\nu}{10^{-9}\eV^2/\MeV}.$$
The gradient is evaluated around the resonance
point  (for a more accurate approximation see~\cite{QVO})
where the density is $N_e \sim
\Delta m^2/G_F E_\nu$: this corresponds to the outer part of the sun
where the profile density deviates from the simple
exponential approximation,
$N_e \propto \exp (-10.54\,r/R_{\rm sun})$.

\smallskip

Before concluding,
we recall that
an accurate treatment of CNO and $pep$ neutrinos
would require to know the actual performance
of the near-future detectors.
However, the sensitivity of the determination of
$\Delta m^2$ through Beryllium neutrinos
would be not  substantially affected,
even in the most pessimistic case.
Instead, in the most optimistic
case, $pep$ neutrinos could give an additional
modulated signal with frequency
$k_{pep}\approx 0.6 k$ and amplitude $a_{pep}\sim a/10,$
referred to those of Beryllium neutrinos, eq.\eq{ka}.
Since the $pep$ flux is well known,
almost as the $pp$ flux,
$a_{pep}$ could yield information on $\theta$.

\subsection*{VO}
Vacuum oscillations offer the same seasonal signal as discussed in the previous section.
Solar matter effects can now be neglected, so that
the energy-averaged survival probability is
$\md{P_{ee}}=1-\frac{1}{2}\sin^2 2\theta$.
The number of seasonal vacuum oscillations $N_{\rm osc}$ encountered during one semester is smaller,
so that the accuracy in $\Delta m^2$ is somewhat worse (see eq.\eq{dm}),
but remains much better than sub-MeV capabilities.
A measurement of the $pp$ rate would give a solar-model independent information on the mixing angle.
Some vacuum oscillation `solutions', usually named ``Just So$^2$'',
that give poor fits of existing data\footnote{We get $\Delta \chi^2\approx 15$, 
which is in agreement with the analyses
in~\cite{standardFit}, except the second one. 
In these `solutions', $P_{ee}$ deviates from unity
only for low energy neutrinos:
if the $3$ standard-deviation evidence from SK and SNO against this possibility is correct,
new SNO data will soon make it even more disfavoured.}
present characteristic spectral distortions in $pp$ neutrinos~\cite{JS2}.
Our point N gives a rather good fit to the data.
In fact, the Beryllium line is affected by seasonal effects 
(while the other fluxes are less affected
due to their larger energy spread)
in a way that very strongly depends on $\Delta m^2$:
choosing the appropriate value one can 
obtain the experimentally preferred value of the Beryllium rate.

\subsection*{SMA}
The SMA region is
strongly disfavoured by existing
experiments and has a low goodness-of-fit.
The reason is that the SMA oscillations
that fit the measured rates imply
a survival probability $P_{ee}(E_\nu)$ in
conflict with the SK spectral data.
This conflict can be seen in pre-SNO
fits performed by the SK collaboration~\cite{SKsun} and
has become sharper after SNO.
Our SMA point, F, is not
the `best' current SMA solution (that has
$\Delta \chi^2\approx 14$, similar
to~\cite{standardFit,statistica})
but a representative point of this region.
In view of this situation, it is difficult
to study seriously how well new experiments can
measure $\Delta m^2$ and $\theta$ in the SMA region.

The most characteristic
feature of the SMA region is that
\begin{itemize}
\item[1.] the neutrino rate at Borexino will be strongly suppressed
(almost down to the background level).
\end{itemize}
Depending on the actual SMA oscillation, this evidence for SMA
can be  stronger than the present evidence against SMA.
In this situation one could doubt that 
Borexino will be able to detect solar neutrinos at all.
Therefore we also assume that
\begin{itemize}
\item[2.] SNO will see a distortion of the
spectrum, and will contradict the SK spectral
data.\footnote{
The first SNO spectral data indicate that this is not the case.
The final SNO energy
spectra are expected to be as significant as those of SK
(SK will have more statistics, but the measurable recoil electron
energy in SNO is
more strongly correlated with the neutrino energy than in SK).
In general, {\em we have no reason to suspect that any solar neutrino data be wrong}.
However, if SMA were
the solution of the solar neutrino anomaly,
some of the present data that strongly disfavour SMA
should be wrong.
It is not true that $\Delta \chi^2\approx 14$
is less statistically significant than a direct
$\sqrt{14}\sim 4$ standard deviations evidence,
because such a $\Delta\chi^2$ is obtained by merging several data
or because oscillations have more than one free parameter.
In fact, 
Borexino should measure a Beryllium rate 
$4$ standard deviations
lower than the one predicted by the current best-fit LMA solution,
in order to make again SMA the `best' fit solution.}
\end{itemize}
Assumptions 1 and 2 imply a very bad goodness-of-fit.
A good SMA fit could be obtained if the
SK collaboration would commit hara-kiri, admitting:
\begin{itemize}
\item[3.]
serious faults in the SK solar neutrino results.
\end{itemize}
In this case, we would be authorized to drop the
the SK spectral data from the global fit.
Under these three assumptions,
we obtain the SMA future fit in fig.\fig{future}.
Borexino data select the SMA region over the other ones,
but without favouring any particular
corner there; this selection is done by the other data, that
prefer the region of {\em large} $\theta$.
We can consider a different possibility:
if the CC rate measured at SNO is wrong
(so that we drop it from the $\chi^2$),
 but the SK spectral data are correct,
the SMA range with {\em small} $\theta$ would be selected.
Incidentally, this tension between the preferred
$\theta$ range reflects once again the
tension between the data, once we assume SMA.

In any case, if something like this were to
happen, we would certainly need new experiments to confirm it.
The $pp$ rate should be consistent with no oscillation, because
SMA predicts $P_{ee}\approx 1$ at very low energy.
In order to make the solar neutrino anomaly credible, we would
need to have a spectral measurement at
low energy, aimed at revealing the sharp SMA 
transition from $P_{ee}\approx 1$ to $P_{ee}\approx 0,$
for energies around the value in eq.\eq{noMSW}.

\begin{figure}[t]
$$
\includegraphics[width=80mm]{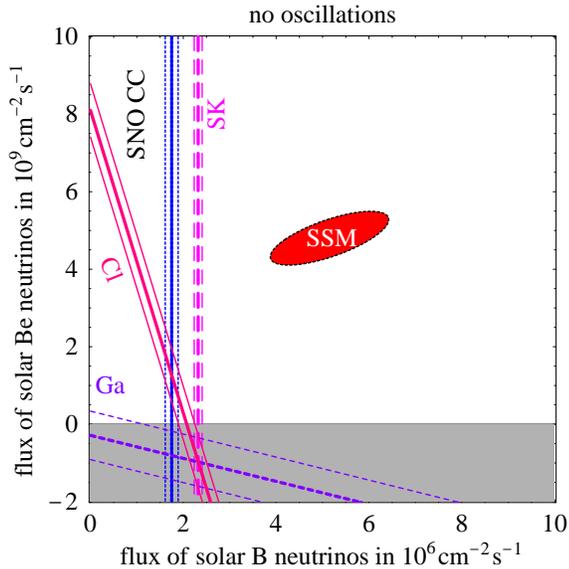} $$
  \caption[]{\em Values of the Boron and Beryllium fluxes 
required at $1\sigma$ 
by the four measured solar neutrino rates in absence of
oscillations, compared to the $68\%$ C.L.\ ellipse of solar model predictions.
\label{fig:noOsc}}
\end{figure}

\subsection*{No oscillations}
The hypothesis of no oscillations can
be reconciled with the data
if the Boron and Beryllium fluxes are very different
(smaller) from solar model predictions, as shown in fig.\fig{noOsc};
see~\cite{SSMindep} and ref.s therein.
Even in this case,
the rates measured by the various experiments
are not well compatible between them.
But, as shown in fig.\fig{noOsc},
there is a significant (accidental?), partial
overlap of the experimental values.\footnote{If GNO
will reduce the error (or the central value)
of the Gallium rate, the crossing in fig.\fig{noOsc}
will happen
at negative unphysical values of the Beryllium flux.
The discrepancy
between the SK and SNO bands is the well known
solar-model-independent evidence
for appearance of $\nu_{\mu,\tau}$ neutrinos.}
This is why solar-model
independent considerations cannot strongly disfavour
the no-oscillations hypothesis --- whereas generically 
such accident does not happen 
and useful solar-model-independent information can be 
deduced already from existing data~\cite{SSMindep}.

The zero Beryllium flux required by no
oscillations is disfavoured by helioseismology~\cite{helio},
by simple physics considerations, and by
recent results. Indeed,
solar models~\cite{BP} have been recently validated by two facts.
The determination of $S_{17}$
(strength of the
$p {}^7{\rm Be}\to {}^8{\rm B}\gamma$ reaction)
has been improved~\cite{ISOLDE,Ham},
and as a consequence the correlation
between the flux of Boron and Beryllium neutrinos, visible in fig.\fig{noOsc},
will tighten (both neutrino fluxes are controlled by the
$\alpha {}^3 {\rm He}\to {}^7 {\rm Be}\gamma$
reactions; thence the relative
$S_{34}$-factor remains  the dominant, common 
source of uncertainty).
Furthermore, the SNO/SK
measurement of the Boron flux~\cite{SNOsun},
is in agreement with the solar-model prediction,
within the $\sim 20$ \% error.
These two facts suggest that the calculated
Beryllium flux should not be {\em too wrong}.\footnote{Conversely,
nuclear data \cite{Ham}
and existing measurements \cite{SNOsun}
tend to suggest that $S_{34}$
(and thence the Beryllium flux) is on the
large side of the predicted value.}
Certainly, the solar model independent
considerations will become cogent after the
Borexino measurement of the Beryllium flux.

\smallskip

Since the prediction of the $pp$ flux
is very accurate~\cite{BP}
$\Phi_{pp} = (5.96\pm 1\%) 10^{10}/\hbox{cm}^2\hbox{s},$
its determinations will also have an important
impact on these analyses.
Indeed, a well known simple (though, less accurate) argument
leads to the same value of the flux.
We assume that the
solar energy comes from nuclear reactions
that reach completion,  and that the sun
is essentially static over the time
employed by photons to reach the surface.
The total luminosity of the
sun, $K_\odot = 8.53\cdot 10^{11}$ MeV cm$^{-2}$ s$^{-1}$
at the earth, determines its total neutrino luminosity as
$K_\odot =  \sum_\alpha \left({Q}/{2} -
\langle E_{\nu_\alpha} \rangle \right)\Phi_\alpha;$
$Q = 26.73$ MeV is the energy released in the reaction $4p  + 2e \rightarrow
\,^4\!{\rm He} + 2\nu_e,$ and
the sum extends over the various components
($\alpha =  pp, pep,
\rm ^7\!Be,^{13}\!N,^{15}\!O,{}^{17}\!F,{}^8\!B,h\hbox{$ep$}$).
Neglecting $\langle E_{\nu_\alpha} \rangle$
and considering only the dominant $pp$ flux, one obtains
$\Phi_{pp}\approx 2K_\odot/Q =   6.4\cdot 10^{10}/\hbox{cm}^2\hbox{s}$,
that is only $7\%$ off.
Therefore, a measurement of the $pp$ rate, even
with modest accuracy, will provide
strong solar-model-indepen\-dent evidence
against (or for) the extreme possibility of no oscillations.

\subsection*{Sterile neutrinos}\label{sterile}
The LSND anomaly can be used to argue
for additional sterile neutrino(s);
however, the evidence against such neutrino(s)
is now stronger than the evidence for them.
For example, the `2+2' best solution
found in~\cite{GMP} where LSND is explained
has a worse global fit than the 3-neutrino
solution where it is not
(and consequently than the best `3+1' solutions,
where the sterile neutrino is of little
use for LSND~\cite{3+1}). Indeed, the relevant
global $\chi^2$ is:
$$\chi^2_{\rm global} =
\chi^2_{\rm sun} + \chi^2_{\rm atm} + \chi^2_{\rm LSND}.$$
Using one sterile neutrino, the best fit gives~\cite{GMP}
$$\chi^2_{\rm global}= 40+36+0=76\qquad \hbox{with 9 parameters.}$$
Without using the sterile neutrino, the $\chi^2$ is~\cite{GMP}
$$\chi^2_{\rm global} = 37+29+10=76\qquad \hbox{with 5 parameters.}$$
CP-violating phases are not counted as relevant parameters.
The $\chi^2_{\rm global}$ of~\cite{GMP}
does not take into account data from SNO; from Karmen;
from nucleosynthesis; from unpublished
atmospheric SK data about the total $\pi^0$ rate
(and recent K2K cross section determinations) and about
their zenith angle dependence~\cite{pi0}.
% as well as the full data from MACRO~\cite{Macro}.
Each one of these data further disfavours
the best-fit `2+2' solution of~\cite{GMP},
which has a large sterile component in atmospheric oscillations. From
a theoretical point of view, `2+2' schemes need a very special
arrangement of  mixing angles.
This pattern can be obtained from a `pseudo-Dirac'
mass matrix, which can be  justified by
a broken U(1) $L_\mu - L_\tau-L_s$ symmetry,
and implies quasi-maximal mixing
in atmospheric oscillations.

If MiniBoone will confirm LSND, it will be interesting to consider
solar oscillations into a  mixed sterile/active neutrino.
What would be the impact on the present analysis?
The near-future prospects concerning the
determination of the solar parameters $\theta$
and $\Delta m^2$ are almost unchanged.
This is an exact statement in the
LMA region, since KamLAND cannot distinguish
if reactor $\bar{\nu}_e$ disappear into
active or sterile antineutrinos.
The most relevant new issue is that
solar oscillations would depend on a
third parameter (other than $\Delta m^2$ and $\theta$),
that quantifies the sterile component in solar oscillations
(for a precise definition, see e.g.~\cite{GMP}).
In this case, it would be interesting to supplement
the NC/CC measurement of the Boron flux performed at SNO and SK
with a NC/CC measurement of the $pp$ flux,
because it is accurately predicted by solar models.
Furthermore, it could be convenient to obtain a CC
measurement of the Beryllium flux, which could be combined
with the result of Borexino (KamLAND).

% Given the fact that either solar or
% atmospheric oscillations have to be significantly
% affected by `2+2' sterile oscillations,
% $\nu_\tau$ appearance at long-baseline experiments
% would play a crucial r\^ole.

\section{Summary}
In the near future KamLAND or Borexino should identify the true
solution of the solar neutrino problem, if it is due to oscillations.
Depending on the actual value of the oscillation
parameters $\Delta m^2$ and $\theta$, the future
situation will be very different, and will
correspondingly require different new experiments.
We summarize the various possibilities below
(see the text for a more detailed discussion):
\begin{itemize}
\item LMA. KamLAND or sub-KamLAND will measure $\Delta m^2$ with
few {\em per-cent} accuracy.
Even the mixing angle $\theta$ can be 
determined reasonably well by reactor experiments; 
it will be a real challenge for sub-MeV experiment 
to improve on these measurements.

\item LOW. Borexino will see
day/ni\-ght effects and measure $\Delta m^2$ with $10\%$ accuracy.
A measurement of the $pp$ rate would
be useful for determining $\theta$ and a measurement
of day/night effects in $pp$ neutrinos could
help in determining $\Delta m^2$.

\item LOW/QVO boundary.
No unmistakable oscillation effect
will be found, but all other solutions will be excluded.
A measurement of day/night effects
(that are largest for the lowest-energy $pp$ neutrinos)
would be crucial.

\item QVO and VO. Borexino will see seasonal
effects and measure $\Delta m^2$ with few {\em per-mille} accuracy,
that can be improved with more statistics.
A measurement of the $pp$ rate would be useful.
\end{itemize}
We comment also on strongly
disfavoured possibilities, that
could strike back again,
if near-future experiments will contradict some
combination of established data:
\begin{itemize}
\item SMA. Will become again the best solution
if Borexino finds almost no solar neutrinos
(because all Beryllium $\nu_e$ get converted);
one has to assume e.g.\ that the SK spectral data are not correct.
A spectral measurement around $E_\nu\sim \MeV$
would provide a crucial signal.

\item No oscillations.
Could become the `best' solution if
Borexino finds no solar neutrinos
(because the Beryllium flux is much smaller
than what solar models and helioseismology tell us ---
this possibility looks very remote) and if the CC/NC rate measured 
at SNO/SK is incorrect.
The measurement of the $pp$ rate would be
of essential importance.

\item Just So$^2$.
Will become the best solution if the CC/NC rate measured 
at SNO/SK will change (contradicting present data) indicating no oscillation,
and if Borexino will find a somewhat suppressed flux of  Beryllium neutrinos.
A measurement of the $pp$ spectrum
would be crucial.

\end{itemize}
Near-future experiments will
allow us to deduce the $pp$ flux
from GNO data, without assuming
oscillations or using solar model predictions,
with $\circa{<} 15\%$ uncertainty.

\paragraph{Acknowledgments}
We thank E.\ Bellotti,  C.\ Cattadori, N.\ Ferrari 
A.\ de Gouv\^ea, A. Ianni, M.\ Junker  and S.\ Sch\"onert
for useful discussions.

\appendix

\section{A seasonal $\chi^2$ for Borexino}
The standard procedure
employs a certain number $ N_{\rm bins}$ of seasonal
bins and a $\chi^2$ analogous to the one
defined in eq.\eq{chiqstandard} for the day/night analysis.
As is apparent from fig.\fig{seasonal},
the appropriate number of seasonal bins is proportional to
the unknown value of $\Delta m^2$, and
many seasonal bins are necessary when $\Delta m^2
\circa{<} 10^{-8}\eV^2$ because $N_{\rm osc}$ is large.
However, eq.\eq{chiqstandard} cannot be applied if
the number of events in each bin
is not much larger than 1.
In this situation, one should not employ Gaussian statistics;
instead, the likelihood of the given measurement, given the
oscillation parameters reads:
$$e^{-{\chi^2}/{2}}=\prod_{i=1}^{N_{\rm bins}}
\frac{\epsilon_i^{n_i}}{n_i!} e^{-\epsilon_i} ,$$
where the individual factors are the Poisson
probabilities of having $n_i$ observed events
in the $i^{th}$ bin, in which $\epsilon_i$ events
are expected.
This  $\chi^2$ function can be written as:
$$\chi^2=2N_{\rm th} -2\sum_m  m
\sum_{i_m=1}^{N_m}\log\epsilon_{i_m} +{\rm const,}$$
where
$N_{\rm th} $ denotes the total number of expected events and
the index $i_m$ runs over the $N_m$ bins with
 $m=\{1,2,\ldots\}$ observed events, so that
$N_0+N_1+N_2+\ldots =N_{\rm bins}$.\footnote{The constant
is irrelevant for parameter estimate.
This $\chi^2$ does not provide a useful goodness-of-fit test,
as any $\chi^2$ with too many bins.}

Poisson statistics simplifies in
the limit of a very large number of seasonal bins
(e.g.\ 1 bin {\em per} millisecond), so that $\epsilon_i\ll 1$.
In this situation there are only bins with
0 or 1 measured events, so that $N_0+N_1=N_{\rm bins}$ and
\begin{equation}
\chi^2
= 2 N_{\rm th} -2 \sum_{i=1}^{N_1}\log\epsilon_{i}
\label{eq:chiqnonstandard}
\end{equation}
If $N_{\rm osc}$ were large, this
$\chi^2$ may be useful to perform
a seasonal analysis at Borexino, already
after the first 6 months of data.

%We can define a more powerful $\chi^2$ by noticing
%that Poisson statistics simplifies also
%in the opposite limit of an infinite number
%of bins with a  number $\epsilon_i$ of expected events much smaller than 1.
%All bins will have zero events, except for $N$ bins with one event:
%$${\wp}(0,\epsilon)= 1-\epsilon+{\cal O}(\epsilon^2),\qquad {\wp}(1,\epsilon) =
%\epsilon+{\cal O}(\epsilon^2).$$
%Therefore, writing the predicted number of events at Borexino as a
%function of the Earth-sun distance as $N(L(t))\,dt$ where
%$$N\propto \frac{1}{L^2}
%\bigg[P_{ee}(L) \sigma_{\nu_e e} + (1-P_{ee}(L))
%\sigma_{\nu_{\mu,\tau}}\bigg] + B$$
%($B$ is the background rate)
%and omitting an irrelevant overall
%constant\footnote{This $\chi^2$ with $\infty$ degrees of freedom
%cannot be used for studying the goodness-of-fit.
%In general, a $\chi^2$ test done using too many bins does not provide
%an efficient goodness-of-fit test:
%this is an extreme example of this fact.} we get
%\begin{eqnarray}\label{eq:chiqnonstandard}
%\chi^2(\Delta m^2,\theta) &=& -2\ln \prod_{\ell=0}^\infty
%{\wp}(N_\ell^{\rm exp},N_\ell^{\rm th}) \\
%& =&-2\sum_{i=1}^{N_{\rm exp}} \ln N(L_i)+ 2 N_{\rm th} \nonumber
%\end{eqnarray}
%where the sum is over the $N_{\rm exp}$ detected events,
%$L_i$ is the earth-sun distance when event $i$ is detected,
%and $N_{\rm th} =\int N(L) dL$ is the total number of
%expected events.

We compare our $\chi^2$ with other approaches.
A standard Fourier-transform analysis of Borexino seasonal data
has been suggested in~\cite{Borexino-QVO1}.
Equation\eq{chiqnonstandard} is, essentially, a non-standard type of transform,
performed with respect to the event rate predicted by oscillations,
rather than to the standard Fourier basis of `$\sin$' and `$\cos$'
functions.
This non standard choice minimizes the uncertainty
on $\theta$ and $\Delta m^2$.
A similar idea has been proposed in~\cite{Baltz}
(where a rough approximation to the
spectrum of Beryllium neutrinos
has been employed) but its
statistical meaning is not clear.
Systematic and theorethical uncertainties can be easily taken into account,
by writing the expected number of events as a function of `nuisance' unknown parameters,
in analogy with eq.\eq{chiqstandard}.
Our definition of the
$\chi^2$ is essentially the same as those
employed in analyses of the SN~1987A data~\cite{SN1987A}.

%   The first term in the $\chi^2$ tells
%   the value of $\Delta m^2$, the second term
%   tells the value of $\theta$.
%   It is useful to see how this $\chi^2$ is equivalent
%   to the standard one in simple cases.
%   Consider an $L$-independent $R(L)$ (no seasonal variations),
%   so that one only wants to deduce the true rate $N_{\rm th}$
%   from the measured
%   number of events $N_{\rm exp}$.
%   The $\chi^2$ in eq.\eq{chiqnonstandard} simplifies to
%   $\chi^2 = 2(N_{\rm th} - N_{\rm exp}\ln N_{\rm th})$.
%   Around $N_{\rm th}\approx N_{\rm exp}$ it agrees with the Gaussian form
%   $$\chi^2(N_{\rm th}\approx N_{\rm exp})
%   = \hbox{cte}+(N_{\rm exp}-N_{\rm th})^2/N_{\rm exp}$$
%   i.e. $N_{\rm th} = N_{\rm exp} \pm \sqrt{N_{\rm exp}}$.

\section{Details of the computation}
Unfortunately, fitting solar neutrino data is a subtle issue:
In order to perform inferences on the parameters of oscillation,
one has to merge together
many pieces of data and information (nuclear physics, solar models,
matter effects in the sun and in the earth,
various experiments,~\ldots).
We used information from many
papers~\cite{ClSun,GaSun,SAGEsun,SKsun,GNOsun,SNOsun,BP,CHOOZ,
LisiChiq,QVO,MSW,Parke,BahcallWWW,PREM,ErroriSpettrali,SNOsigma,ExpResolution}.
It is briefer to list here what
is {\em not} included in our fit.
Most of the other global fits have similar shortcomings.

The spectrum of recoil electrons
in SK and SNO is computed in the simplest approximation
(e.g.\ neither one-loop effects
nor photon emission~\cite{Passera} are included).
One-loop corrections to the MSW effect are neglected.
Seasonal Gallex/GNO and SAGE data are not included.
The treatment of solar matter effects in the QVO
region is not as precise as in~\cite{QVO}.
When computing confidence levels, we
approximate $e^{-\chi^2/2}$ with a Gaussian
function of $\Delta m^2$ and $\theta$:
correct frequentistic and Bayesian
analyses~\cite{statistica} do not give a
significantly different result.
All above issues do not have significant effects.
Alternative possible definitions of the $\chi^2$
give $\Delta\chi^2$ values similar to those quoted in table~1.
Note that the first digit of the $\Delta\chi^2$ is significant.

Our fit of SK data is based on table III of~\cite{SKsun}.
It only allows us to reproduce the total rate
and the total day/night asymmetry (quoted in~\cite{SKsun})
with $\sim 1\sigma$ accuracy.
This small discrepancy is presumably due to the fact that
the $\chi^2$ used by the SK collaboration includes a proper
treatment of data about the background,
so that the total rate is not the sum of the
rates in each energy bin:
the bins with higher energy are relatively
more important, since they have less background.
%  The errors on the single bins quoted in table III of~\cite{SKsun}
%  do not allow to reproduce the error on the total rate,
%  quoted in eq.~(XYX) of~\cite{SKsun}. bla bla

\footnotesize
% \begin{multicols}{2}

%\end{multicols}

\end{document}